\documentstyle[11pt,paspconf,epsf]{article}

\begin{document}

\title{Tidally--induced warps in protostellar discs }

\author{C. TERQUEM}
\affil{Lick Observatory, University of California \\
Santa Cruz, CA~95064, USA}

\author{J. C. B. PAPALOIZOU and R. P. NELSON}
\affil{Astronomy Unit, Queen Mary \& Westfield College \\
Mile End Road, London~E1~4NS, UK}

\begin{abstract}
We review results on the dynamics of warped gaseous discs.  We
consider tidal perturbation of a Keplerian disc by a companion star
orbiting in a plane inclined to the disc. The perturbation induces the
precession of the disc, and thus of any jet it could drive. In some
conditions the precession rate is uniform, and as a result the disc
settles into a warp mode. The tidal torque also leads to the
truncation of the disc, to the evolution of the inclination angle (not
necessarily towards alignment of the disc and orbital planes) and to a
transport of angular momentum in the disc.  We note that the spectral
energy distribution of such a warped disc is different from that of a
flat disc. We conclude by listing observational effects of warps in
protostellar discs.
\end{abstract}



\section{Introduction}

\subsection{Observations of pre--main sequence binary systems}

In the course of the past few years, accretion discs around low--mass
pre--main sequence stars known as T~Tauri have been directly imaged by
the HST. As expected, they appear to be common (Stauffer et al.~1994).

Furthermore, most T~Tauri stars are observed to be in multiple systems
(see Mathieu~1994 and references therein). In some of these systems,
one or even two circumstellar discs are resolved or inferred from the
spectral energy distribution of the stars.  This is the case for
UY~Aurigae (Duvert et al. 1998, Close et al. 1998), T~Tau (Akeson et
al. 1998), HK~Tau (Stapelfeldt et al. 1998, Koresko~1998), although in
that case it is not certain whether the two stars are bound or not,
and GG~Tau (Roddier et al. 1996). Very recently, it has been announced
that two circumstellar discs made of dust have been imaged by the VLA
in a binary located in Taurus (L.~F.~Rodriguez et al. 1998, press
release). The binary separation is about 45~AU, and the radius of each
disc is on the order of 10~AU.

There are indications that the plane of at least one of the
circumstellar discs and that of the orbit may not necessarily be
aligned.  The most striking evidence for such non--coplanarity is
given by HST and adaptive optics images of HK~Tau (Stapelfeldt et
al. 1998, Koresko~1998).  The binary system T~Tau is also very likely
to be non--coplanar, because two bipolar outflows of very different
orientations originate from this system (B\"ohm~\& Solf~1994).  Since
it is unlikely that they are both ejected by the same star, each of
them is more probably associated with a different member of the
binary. Furthermore, jets are usually thought to be ejected
perpendicularly to the disc from which they originate. These
observations then suggest that discs and orbit are not coplanar in
this system.  More generally, observations of molecular outflows in
star forming regions commonly show several jets of different
orientations emanating from an unresolved region the extent of which
can be as small as a few hundreds astronomical units (Davis et
al. 1994).

\subsection{Tidal interactions in binary systems}

There is evidence that the sizes of circumstellar discs contained
within binary systems are correlated with the binary separation
(Osterloh~\& Beckwith~1995, Jensen et al. 1996). This suggests that
binary companions are responsible for limiting the sizes of the discs
through tidal truncation (see, e.g., Papaloizou~\& Pringle~1977,
Paczy\'nski~1977, Artymowicz~\& Lubow~1994).

The tidal effect of an orbiting body on a differentially rotating disc
has been well studied in the context of planetary rings (Goldreich~\&
Tremaine~1978), planet formation and interacting binary stars (see
Lin~\& Papaloizou~1993 and references therein).  In these studies,
disc and orbit are usually taken to be coplanar, so that the only
waves excited in the disc by the perturbing companion are density
waves.

However, as mentioned above, the plane of at least one of the
circumstellar discs may not be aligned with that of the orbit. If this
is the case, the companion's perturbation excites both density and
bending waves in the disc and induces it to warp.  Here we will focus
on the effects of this warping and the propagation of bending waves.
We note that when the perturbation is linear, the effects of density
waves may be superimposed on those of bending waves.

The discs we consider here are gaseous. The work we present should
then be contrasted with studies of warping of purely viscous discs in
which pressure and self--gravity are ignored (Bardeen~\&
Petterson~1975, Katz~1980, Steiman--Cameron~\& Durisen~1988,
Pringle~1996). In a purely viscous disc, evolution occurs only through
viscous diffusion, whereas in a gaseous disc pressure effects
manifesting themselves through bending waves can control the disc
evolution (see also the chapter by Nelson et al. in this book). When
these waves propagate, they do so on a timescale comparable to the
sound crossing time (Papaloizou~\& Lin~1995). Since this is much
shorter than the viscous diffusion timescale, communication through
the disc occurs as a result of the propagation of these waves.  Even
when the waves are damped before reaching the disc boundary, a
diffusion coefficient for warps that is much larger than that produced
by the kinematic viscosity may occur because of pressure effects
(Papaloizou~\& Pringle~1983).

\section{Perturbing potential}

\begin{figure}
\plotone{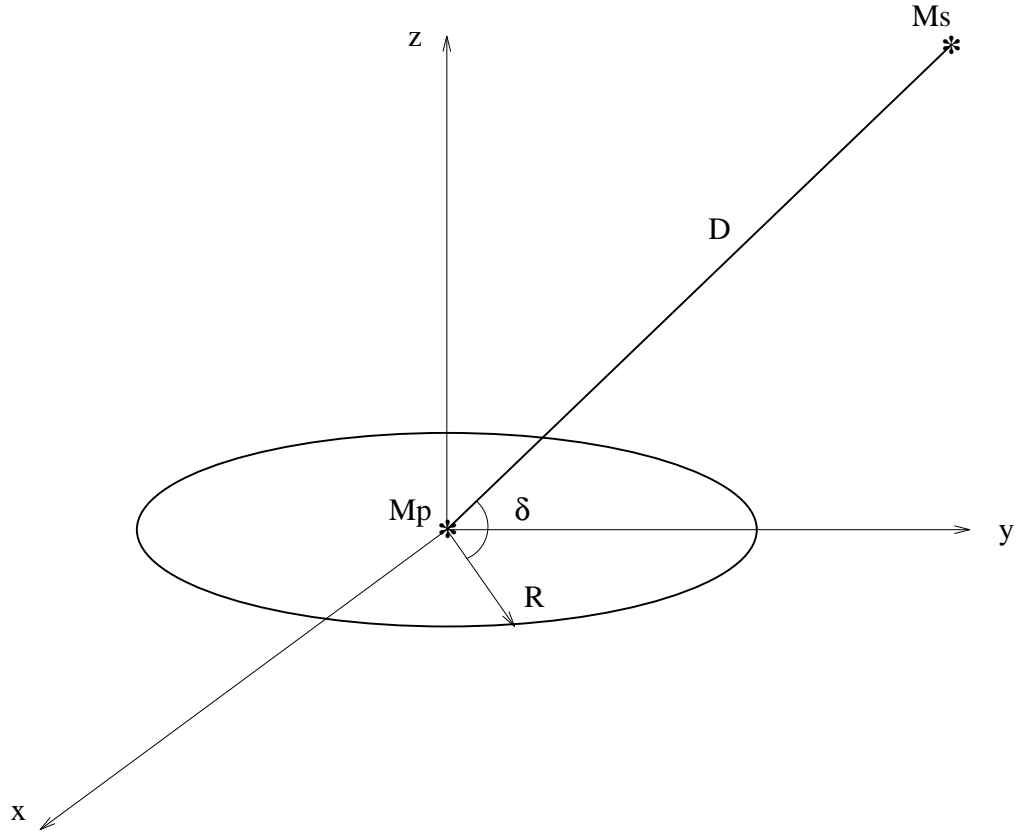}
\caption[]{Sketch of the binary system.}
\label{fig1}
\end{figure}

We consider a binary system in which the primary has a mass $M_p$ and
the secondary has a mass $M_s$.  The binary orbit is assumed circular
with radius $D.$ We suppose that the primary is surrounded by a disc
of radius $R \ll D$ with negligible mass so that precession of the
orbital plane can be neglected. The secondary describes a prograde
Keplerian orbit about the primary with angular velocity $\omega$. We
define a Cartesian coordinate system $(x,y,z)$ centered on the primary
star, where the $z$--axis is normal to the initial disc mid--plane. We
also define the associated cylindrical polar coordinates $(r, \varphi,
z)$.  The orbit of the secondary star is in a plane which has an
initial inclination angle $\delta$ with respect to the $(x,y)$ plane.
This situation is illustrated in Figure~\ref{fig1}.

We are interested in disc warps which are excited by terms in the
perturbing potential which are odd in $z$ and have azimuthal mode
number $m=1$ when a Fourier analysis in $\varphi$ is carried
out. There are three terms in the perturbing potential with the
required form. One of them is secular (time independent), while the
two others, with frequency $2 \omega$ and $-2 \omega$, are prograde
and retrograde, respectively (see Papaloizou~\& Terquem~1995 for more
details).

\section{Precession of warped discs}

\subsection{Theory}

Hunter~\& Toomre~(1969) showed that an isolated self--gravitating disc
subject to a vertical displacement generally precesses differentially,
and thus cannot sustain a warped configuration.  However, differential
precession is prevented by gravitational torques from the distorted
disc itself if the disc has a sharp edge.  A similar process can occur
if the disc orbits in the external potential due to a companion
(Hunter~\& Toomre~1969) or a flattened halo whose equatorial plane is
misaligned with the disc plane (Toomre~1983, Sparke~1984, Sparke~\&
Casertano~1988).

Papaloizou~\& Terquem~(1995) have shown that {\it radial pressure
forces are also able to smooth out differential precession} in a non
self--gravitating inviscid accretion disc.  This process can be
effective if the sound crossing time through the disc is much smaller
than the precession period. {\it The condition for rigid body
precession is then $H/r > \left| \omega_p \right| / \Omega_o$}, where
$\omega_p$ is the (uniform) precession frequency, $H$ is the disc
semi--thickness and $\Omega_o$ is the angular velocity at the disc
outer edge.  When this condition is satisfied, bending waves are able
to propagate through the disc sufficiently fast so that the different
parts of the disc can ``communicate'' with each other and adjust their
precession rate to a constant value. This also happens when viscosity
is present, but the communication becomes diffusive rather than
wave--like when the Shakura~\& Sunyaev~(1973) viscosity parameter
$\alpha$ significantly exceeds $H/r$ (Papaloizou~\&
Pringle~1983). However, in protostellar discs, we expect $\alpha < H/r$ 
so that, assuming the effective viscous stress
tensor is isotropic, disc communication is governed by bending waves.

The disc precesses because, in a non--rotating frame, the component of
the secular torque along the direction of the line of nodes is
non--zero. The precession frequency $\omega_p$ can then be calculated
from the condition that, in a rotating frame, this component of the
torque is balanced by the Coriolis torque. This leads to
(Papaloizou~\& Terquem~1995; see also Kuijken~1991 in the context of
galactic discs):

\begin{equation}
\frac{\omega_p}{\Omega_o} = - \frac{3}{4} \frac{M_s}{M_s+M_p} \left(
\frac{\omega}{\Omega_o} \right)^2 \cos \delta \int_{r_{in}}^R
\frac{\Sigma}{\left(\Omega/\Omega_o\right)^2} dr \left/
\int_{r_{in}}^R \frac{\Sigma}{\Omega/\Omega_o} dr \right. ,
\label{prec1}
\end{equation}

\noindent where $r_{in}$ is the disc inner radius, $\Omega$ is the
angular velocity in the disc, and $\Sigma$ is the disc surface mass
density.

Although the above assumes a circular orbit, an eccentric binary orbit
can be considered by replacing $D$ by the semi-major axis and
multiplying the precession frequency by $(1-e^2)^{-3/2},$ with $e$
being the eccentricity.

The analysis used to derive equation~(\ref{prec1}) is accurate only
when $\left| \omega_p \right| / \Omega_o $ is smaller than the
maximum of $H^2/r^2$ and $\alpha$.

We can approximate $\omega_p$ by assuming that $\Sigma$ is constant
throughout the disc and that the rotation is Keplerian, so that $\Omega
= \sqrt{GM_p/r^3}$.  Equation~(\ref{prec1}) can then be written in the
form:

\begin{equation}
\omega_p = - \frac{15}{32} \frac{M_s}{M_p} \left( \frac{R}{D}
\right)^3 \cos \delta \sqrt{\frac{G M_p}{R^3}}.
\label{prec}
\end{equation}

We note that even though $\omega_p \propto \cos \delta$, the amplitude
of the precessional motion vanishes when $\delta =0$.

These results are completely supported by the non--linear,
three--dimensional hydrodynamic simulations of the tidal perturbation
of accretion discs in non--coplanar binary systems performed by
Larwood~et al.~(1996), using a SPH code (see also the chapter by
Nelson et al. in this book).  They found that the disc tends to
precess approximately as a rigid body if it is not too thin, and the
precession period they derived agree well with the linear estimate
given above.  These simulations also show that extremely thin discs
are severely disrupted by differential precession.  For a binary mass
ratio around unity, $D/R$ in the range 3 to 4, and $\alpha \sim 0.03$
corresponding to the dissipation in the code used here, the crossover
between obtaining a warped disc structure and disc disruption appears
to occur for values of $H/r \sim 0.033$.

We also point out that, according to these simulations, {\it tidal
truncation operates effectively when the disc and the binary orbit are
not coplanar}, being only marginally affected by the lack of
coplanarity.  This is consistent with the observations of HK~Tau
(Stapelfeldt et al. 1998, Koresko~1998).

\subsection{Observational tests: precessing jets}

Observations of molecular outflows in star--forming regions show in
some cases knots forming a helical pattern or 'wiggling' which can be
interpreted as being the result of the precession of the jet (see,
e.g., Bally~\& Devine~1994, Eisl\"offel et al. 1996, Davis et
al. 1997, Mundt~\& Eisl\"offel~1998). Assuming that such precession is
caused by tidal interaction between the disc from which the outflow
originates and a companion star in a a non--coplanar orbit, Terquem et
al. (1998) have estimated the separation of the binary for several
systems. The numbers they found are characteristic of those expected
for pre--main sequence stars.

Also, as mentioned in the introduction, jets with very different
orientations are commonly observed in star--forming regions, and it is
generally assumed that they originate from a binary where the discs
are misaligned. For some systems, a binary has actually been
resolved. It is the case for T~Tau, which is a binary where a
circumstellar disc has been resolved (Akeson et al. 1998) and from
which two almost perpendicular well collimated outflows originate
(B\"ohm~\& Solf~1994). It is also the case for HH~24, which is a
hierarchical system with 4 or even 5 stars and several outflows
(Eisl\"offel, personnel communication). For these systems, Terquem et
al. (1998) have evaluated the precession frequency and given an
estimate of the length--scale over which the outflows should 'wiggle'
as a result of this precessional motion.

\section{Warp mode and disc deformation}

When the disc can precess as a rigid body under the action of the
tidal force, it settles into a discrete bending mode (representing a
warp) which is referred to as the modified tilt mode because in the
limit that the external potential is spherically symmetric it reduces
to the trivial rigid tilt mode. The difference in shape and frequency
between the rigid and modified tilt modes is due to the fact that, in
a non--spherically symmetric potential, the disc has to bend to alter
the precession frequency at each radius so that the rate is everywhere
the same (Sparke~\& Casertano~1988).  This asymptotic state is
possible because bending waves transport away the energy associated
with the transient response (Toomre~1983).  The timescale for settling
to the warp mode is then given by the characteristic time for bending
waves to propagate through the disc but the process may be slowed down
in the low surface density regions near the outer edge.

When pressure is present, the waves may be reflected from the edge
before they attain arbitrary short wavelength there. Some dissipation
is then needed for the disc to settle into a state of near rigid
precession. This can be provided by disc viscosity which is least
effective on a global disturbance such as near rigid body precession.
Other transient disturbances would have shorter wavelengths and
accordingly would be expected to dissipate significantly more
rapidly. In this context note that a warp mode which deviates only
slightly from a rigid tilt has a wavelength which is much larger than
the disc.  In the simulations of Larwood et al. (1996), of non
self--gravitating inclined discs with pressure and viscosity, the disc
was seen to quickly settle into a state of near rigid body
precession. The time for this to happen was consistent with about
$\sim 1/\alpha$ orbital periods, this being the decay time of short
wavelength bending waves without self--gravity (see Papaloizou~\& Lin
1995).

Since the analysis we have performed and we report in this chapter is
linear, its domain of validity is limited (see Terquem~1998 for a
summary). We are interested in protostellar discs in which we expect
$\alpha < H/r$ (we assume here that the effective viscous stress
tensor is isotropic).  In these conditions, the secular perturbation
produces a tilt the variation of which across the disc can be up to
$H^2/r^2$ in the direction perpendicular to the line of nodes and
$\alpha$ (or $H^2/r^2$ if the disc is inviscid) along the line of
nodes.  Superimposed on this tilt, there is another tilt produced by
the finite frequency perturbations, the variation of which across the
disc can be up to $H/r$.  Such deformations were observed in the SPH
simulations performed by Larwood~\& Papaloizou~(1997), in which
$\alpha > H^2/r^2$.

We note that since in protostellar discs $H/r \sim 0.1$, the variation
of the vertical displacement across the disc due to the finite
frequency perturbations can be as large as about a tenth of the disc
radius while the perturbation remains linear.

\section{Evolution of the inclination angle}

In section 3 we have described the effect of the component of the
secular tidal torque along the direction of the line of nodes. We now
discuss the effect of the component of this torque along the axis
perpendicular to the line of nodes in the disc plane. It induces the
evolution of the inclination angle of the disc plane with respect to
that of the orbit.

Papaloizou~\& Terquem~(1995) have found that {\it this evolution does
not necessarily tend to align the disc plane with that of the
orbit}. The situation were these planes are aligned may indeed be
unstable. 

They have also shown that, in a non--self gravitating disc, this
evolution occurs on a timescale related to the rate at which angular
momentum is transferred from the disc rotation to the orbital motion.
This conclusion might be expected since the disc inclination as a
whole can change only if angular momentum is transferred between
different parts of the disc and then between the disc and the
companion's orbit. In the case we consider here, angular momentum
exchange between the disc rotation and the orbital motion takes place
as a result of the viscous shear stress, which induces a lag between
the response of the disc and the perturbing potential (see next
section).  Therefore we expect the timescale for the evolution of the
inclination angle to be at least the disc viscous timescale.

This was seen in the numerical simulations of Larwood et al. (1996),
in which the disc inclination evolved much more slowly than the
precession rate.

We then conclude that the probability of observing a warped disc, if 
binary systems are not all coplanar when they form, may be significant.

We emphasize that these results only hold if the disc is able to find
a state in which its precesses as a rigid body.  Settling toward a
preferred orientation is indeed much faster if differential precession
is so important that it cannot be controlled by pressure (or
self--gravitating) forces (Steiman--Cameron~\& Durisen~1988).

\section{Angular momentum transport associated with the warp}

So far we have discussed the effect of the components of the secular
torque in the disc plane. We now consider the effect of the component
of the torque (secular or not) parallel to the disc rotation axis
$z$. It does not modify the direction of the disc angular momentum
vector but its modulus, and thus changes the angular momentum content
of the disc.  As a result of this torque, angular momentum is
exchanged between the disc rotation and the orbital motion.

If the disc is inviscid and does not contain any corotation resonance,
the nature of the boundaries determines whether such an exchange takes
place or not, i.e. whether the $z$--component of the tidal torque is
finite or zero (see Lin~\& Papaloizou~1993 for example).  Because of
the conservation of wave action in an inviscid disc, the tidal waves
excited at the outer boundary propagate through the disc with an
increasing amplitude if the disc surface density increases inwards or
is uniform.  It is usually assumed that they become non linear before
reaching the center and are dissipated through interaction with the
background flow. Thus, the inner boundary can be taken to be
dissipative. This introduces a phase lag between the perturber and the
disc response, enabling a net torque to be exerted by the
perturber. This torque is transferred to the disc through dissipation
of the waves at the boundary. Because of the conservation of angular
momentum, the net torque is equal to the difference of angular
momentum flux through the disc boundaries. This flux is constant
(independent of $r$) inside the disc, since there is not dissipation
there. We note that in these conditions only the finite frequency
terms (not the secular term) produce a torque. Whenever the perturber
rotates outside the disc, the torque exerted on the disc is
negative. Through dissipation of the waves, the disc then loses
angular momentum.  Papaloizou~\& Terquem~(1995) have calculated this
torque in the context of pre--main sequence binaries. They found that
$m=1$ bending waves can lead to the accretion of the disc on a
timescale which, if it were written as a viscous timescale, would
correspond to an $\alpha$ parameter smaller than $10^{-4}$. This upper
limit was actually reached only in extreme cases, and the equivalent
$\alpha$ is more likely to be at least one order of magnitude
smaller. We note that of course transport of angular momentum by waves
does not produce an $\alpha$--type disc, since energy is not
dissipated locally but transported by the waves away from the location
where it has been extracted from the disc.
 
We remark that bending waves are more efficient at transporting
angular momentum in the disc than density waves, because they have a
longer wavelength (Papaloizou~\& Lin~1995). For the same reason they
can also affect the disc at smaller radii than density waves.

The situation is different from that described above when a corotation
resonance is present in the disc, since this singularity provides a
location where angular momentum can be absorbed or emitted
(Goldreich~\& Tremaine~1979). However, there is no such resonance in
the cases we consider here.
 
When the disc is viscous, its response is not in phase with the
perturber. A net tidal torque is then exerted on the disc even if the
boundaries are reflective, and the angular momentum flux inside the
disc is not constant, since the perturbed velocities are viscously
dissipated. In that case, both the secular and the finite frequency
perturbing terms produce a torque. Terquem~(1998) has calculated the
tidal torque in this context and has found that it can be comparable
to the horizontal viscous stress acting on the background flow when
the perturbed velocities in the disc are on the order of the sound
speed.  If these velocities remain subsonic, the tidal torque can
exceed the horizontal viscous stress only if the viscous stress tensor
is anisotropic with the parameter $\alpha$ which couples to the
vertical shear being larger than that coupled to the horizontal
shear. We note that, so far, there is no indication on whether these
two parameters should be the same or not. When the perturbed
velocities become supersonic, shocks reduce the amplitude of the
perturbation such that the disc moves back to a state where these
velocities are smaller than the sound speed (Nelson~\&
Papaloizou~1998, see also the chapter by Nelson et al. in this
book). When shocks occur, the tidal torque exerted on the disc may
become larger than the horizontal viscous stress.  Terquem~(1998) also
found that if the waves are reflected at the center, resonances occur
when the frequency of the tidal waves is equal to that of some free
normal global bending mode of the disc.  If such resonances exist,
tidal interactions may then be important even when the binary
separation is large.  Out of resonance, the torque associated with the
secular perturbation is generally much larger than that associated
with the finite frequency perturbations.  As long as the waves are
damped before they reach the center, the torque associated with the
finite frequency perturbations does not depend on the viscosity, in
agreement with theoretical expectation (Goldreich~\& Tremaine~1982).

\section{Effect of the warp on the spectral energy distribution of 
the disc}

The spectral energy distribution of a circumstellar disc can be
significantly affected by a warp
because reprocessing of radiation from the central star by a disc
depends crucially on the disc geometry.

Terquem~\& Bertout~(1993) have shown that T~Tauri stars with tidally
warped circumstellar discs may display far--infrared and
submillimetric flux well in excess of that expected from flat
circumstellar discs. The excess occurs at relatively long wavelength
because the warp affects mostly the outer, low--temperature regions of
the disc, at least when the perturbation is not too severe. 

Terquem~\& Bertout~(1996) produced a broad variety of synthetic
spectral energy distributions of warped discs, and compared them with
actual observations. They found that they could reproduce the spectral
energy distribution of a T~Tauri star with infrared excess (Class~II
source), that similar to GW~Ori with a double-peak or even that of a
Class~I source with not unrealistic, though a bit extreme, warp and
disc parameters. The claim is of course not to provide an explanation
for all these non--standard spectral energy distributions in terms of
a warped disc.  However, it appears that tidal interaction in T~Tauri
binary systems with intermediate separations may play a role in
shaping the spectral energy distributions of these stars.

\section{Observational effects of warps in protostellar discs}

We have already mentioned above that the precession of jets and a
non--standard spectral energy distribution would be an observational
effect of warped discs. As far as precessing jets are concerned, we
hope to be able to compare some predictions of our models with
observations in the near future.

Also, because the variation of the tilt angle along the line of nodes
and along the perpendicular to the line of nodes depend on $\alpha$
and $H/r$, respectively, observations of warped protostellar discs
have the potential to give important information about the physics of
these discs.

Protostellar discs are believed to be rather thick, i.e. $H/r \sim
0.1$. In addition, $\alpha$ is believed to be at most on the order of
$10^{-3}$--$10^{-2}$.  We then expect bending waves to
propagate on a relatively short timescale across the disc.  It may
then be possible to observe some time--dependent phenomena with a
frequency equal to that of these waves, i.e. twice the orbital
frequency. Such a phenomenon may actually have been detected in the
X--ray binary SS~433, under the form of a ``nodding motion''
(Margon~1984).

\acknowledgments We thank the Isaac Newton Institute and programme
organisers for hospitality.

\end{document}